\documentclass[amssymb,prb,twocolumn,floats]{revtex4}
\usepackage{bm}
\usepackage{graphicx}
\begin{document}
\preprint{July 23, 2000}
\title{ Carrier-induced ferromagnetism in p-Zn$_{1-x}$Mn$_{x}$Te}
\author{ D. Ferrand}
\author{ J. Cibert}
\email{cibert@drfmc.ceng.cea.fr}
\author{ A. Wasiela, C. Bourgognon, S. Tatarenko}
\author{G. Fishman}
\thanks{Present address: IEF, UMR 6822 CNRS, Universit\'e Paris-Sud,
F-91405 Orsay C\'edex, France} \affiliation{Laboratoire de
Spectrom\'etrie Physique, CNRS - Universit\'e Joseph-Fourier
Grenoble Bo\^ite Postal 87, F-38402 St Martin d'H\`eres Cedex,
France}
\author{ T. Andrearczyk, J. Jaroszy\'nski, S. Kole\'snik}
\author{T. Dietl}
\email{dietl@ifpan.edu.pl}
\homepage{www.ifpan.edu.pl/SL-2/sl23.htlm} \affiliation{Institute
of Physics, Polish Academy of Sciences,  al.~Lotnik\'ow 32/46,
PL-02-668 Warszawa, Poland}
 \author{ B. Barbara and D. Dufeu}
\affiliation{ Laboratoire de Magn\'etisme Louis N\'eel, CNRS,
Bo\^ite Postal 166X, F-38042 Grenoble-C\'edex, France}
\date{\today}
\begin{abstract}
We present a systematic study of the ferromagnetic transition
induced by the holes in nitrogen doped Zn$_{1-x}$Mn$_{x}$Te
epitaxial layers, with particular emphasis on the values of the
Curie-Weiss temperature as a function of the carrier and spin
concentrations. The data are obtained from thorough analyses of the
results of magnetization, magnetoresistance and spin-dependent Hall
effect measurements. The experimental findings compare favorably,
without adjustable parameters, with the prediction of the
Rudermann-Kittel-Kasuya-Yosida (RKKY) model or its
continuous-medium limit, that is, the Zener model, provided that
the presence of the competing antiferromagnetic spin-spin
superexchange interaction is taken into account, and the complex
structure of the valence band is properly incorporated into the
calculation of the spin susceptibility of the hole liquid. In
general terms, the findings demonstrate how the interplay between
the ferromagnetic RKKY interaction, carrier localization, and
intrinsic antiferromagnetic superexchange affects the ordering
temperature and the saturation value of magnetization in
magnetically and electrostatically disordered systems.
\end{abstract}

\maketitle

\section{INTRODUCTION}

The possibility of controlling ferromagnetic interactions between
the localized spins by the
carriers,\cite{Stor86,Mats98,Diet97,Haur97,Ferr00a} as well as the
demonstration of efficient spin injection into a normal
semiconductor,\cite{Oest99,Fied99,Ohno99} have recently renewed the
interest in diluted magnetic semiconductors (DMS).\cite{Furd88} If
made functional at reasonably high temperature, ferromagnetic
semiconductors would allow one to incorporate spin electronics into
usual electronics, and even path the way to integrated quantum
computers.\cite{Loss98} Up to now, the carrier-induced
ferromagnetism has been observed in lead-salt materials,
Pb$_{1-x-y}$Sn$_y$Mn$_{x}$Te,\cite{Stor86}and in MBE-grown
semiconductors with the zinc-blende structure:
In$_{1-x}$Mn$_{x}$As\cite{Ohno92} and
Ga$_{1-x}$Mn$_{x}$As,\cite{Mats98,Ohno96} as well as in p-doped
Cd$_{1-x}$Mn$_{x}$Te quantum wells\cite{Haur97} and
Zn$_{1-x}$Mn$_{x}$Te epilayers,\cite{Ferr00a} in which the
observation of the onsets of magnetic ordering in the temperature
range between 1 and 3~K corroborated theoretical
predictions.\cite{Diet97} Thus, while promisingly high Curie
temperatures, $T_{\mbox{\small C}}$ up to
110~K,\cite{Mats98,Ohno96} are observed in the GaAs-based
compounds, they are dramatically lower in the II-VI DMS structures
studied so far. However, the II-VI compounds appear as model
materials, in which localized spins and the holes can be introduced
and controlled independently, and modulation-doped heterostructures
are feasible, so that dimensionality effects can be examined.

In this paper, we present the phase diagram of nitrogen-doped
p-Zn$_{1-x}$Mn$_{x}$Te, {\it i.e.}, the dependence of the
Curie-Weiss temperature $T_{\mbox{\small CW}}$ on the Mn content
$x$ and the hole concentration $p$. The values of $T_{\mbox{\small
CW}}$ and $x$ were obtained from magnetization measurements, while
the hole densities were deduced from the Hall resistivity, measured
under such conditions that the spin-dependent component is
negligible, {\it i.e.}, either at room temperature, or at
low-temperature in a high magnetic field. We show that the observed
values of $T_{\mbox{\small CW}}$ are well described by a mean-field
model, in which the hole-mediated exchange interactions are treated
either in terms of Rudermann-Kittel-Kasuya-Yosida (RKKY) coupling
mechanism,\cite{Diet97} or by its continuous-medium limit, {\it
i.e.}, the Zener model,\cite{Diet97,Jung99,Diet00a} when the hole
density is not too high. Accordingly, three parameters govern the
carrier-induced ferromagnetism: the spin-carrier exchange integral
$\beta$, the effective content of magnetic ions $x_{\mbox{\small
eff}}$, and the spin susceptibility of the carrier liquid
$\chi_{\mbox{\small h}}$. These three parameters are well known or
can be readily evaluated for p-Zn$_{1-x}$Mn$_{x}$Te. Our findings
emphasize the importance of taking carefully into account the
competition between ferromagnetic and antiferromagnetic
interactions as well as the complex structure of the valence band
in the calculation of $\chi_{\mbox{\small h}}$.\cite{Diet00a}
Finally, we discuss the case of lower doping, for which the onset
of hole localization is clearly observed. We show that localization
does not perturb significantly the magnitude of Curie-Weiss
temperature but reduces the saturation value of magnetization and
gives raise to slow spin dynamics.

The heavily doped DMS studied here lie in-between the case of
diluted magnetic metals,\cite{Mydo93} where the standard RKKY
theory is applicable, and lightly doped DMS,\cite{Furd88} for which
carriers thermally excited to the band\cite{Pash79} and
interactions between bound magnetic polarons\cite{Wolf96} have been
suggested as agents mediating the ferromagnetic coupling. Our
results make it possible to identify differences and similarities
between ferromagnetic III-V and II-VI material systems.
Furthermore, the findings demonstrate how spin-orbit coupling, the
competing antiferromagnetic interactions, and the presence of
electrostatic disorder affect the carrier-induced ferromagnetism.
The Stoner ferromagnetic instability in disordered conductors at
low temperatures have recently been discussed
theoretically.\cite{Kirk99,Naro00} We hope that our results will
stimulate a similar analysis for the system of localized magnetic
ions coupled ferromagnetically by carriers at the boundary of the
Anderson-Mott localization.

\section{ SAMPLES AND EXPERIMENTAL SET-UP}

Zn$_{1-x}$Mn$_{x}$Te:N layers were grown\cite{Ferr00a} by molecular
beam epitaxy (MBE) on a (001) Cd$_{0.96}$Zn$_{0.04}$Te substrate,
on which a 300 nm CdTe and 200 nm-thick ZnTe buffer layers were
deposited, the latter much thicker than the critical thickness of
the ZnTe/CdTe system. The active layer, Zn$_{1-x}$Mn$_{x}$Te:N,
typically 500 nm thick, was grown at 300$^o$C. The
Zn$_{1-x}$Mn$_{x}$Te layers were deposited either by using the
stoichiometric Zn/Te flux from a ZnTe load and simply adding a Mn
flux during the growth of the active layer (resulting in rather
rough surfaces), or by using a Zn rich flux (adding an excess of Zn
from an additional Zn cell, approximately 50\% of the Zn flux from
the ZnTe cell), which gives a smooth surface. A home-designed
electron cyclotron resonance (ECR) plasma cell served as a source
of atomic nitrogen.\cite{Grun96}

Resistivity and Hall effect measurements were performed between
room temperature and 1.5~K in a magnetic field $H_{\mbox{\small
o}}$ up to 110~kOe, applied in the direction perpendicular to the
film surface. In this geometry, there is no magnetization
corrections to the external magnetic field acting on the carriers,
$B = H_{\mbox{\small o}}$. At the same time, the field acting on
the localized spins is diminished by the demagnetization
correction, $H=H_{\mbox{\small o}}-4\pi M$. For $x=0.015$, the
field produced by the saturated Mn spins is $B_s = 4\pi M_s = 154$
G. Typically, the conductivity and the Hall voltage were measured
on $5\times10$~mm$^{2}$ samples with six gold contacts (without
etching a Hall bar), using a d.c. current between 0.1~nA and 1~mA.
For a Zn$_{0.981}$Mn$_{0.019}$Te:N sample with the highest hole
concentration the measurements were extended down to 100~mK with
the use of the lock-in technique. Magnetization studies were
carried out down to 1.5~K employing either a superconducting
quantum interference device (SQUID) set-up (with the magnetic field
up to 1.3~kOe) or a vibrating sample magnetometer (with the field
up to 20~kOe). In both cases the field was applied in the direction
parallel to the film surface, so that no demagnetization effects
had to be taken into account.

The values of hole and Mn concentrations were obtained from the
studies of the Hall effect and magnetic susceptibility,
respectively, according to procedures described in detail below. A
hole concentration as high as $1.2\times10^{20}$~cm$^{-3}$ was
obtained in the case of ZnTe and Zn$_{0.981}$Mn$_{0.019}$Te
epilayers.\cite{Ferr00a} To our knowledge, this is the largest hole
concentration ever achieved for any II-VI semiconductor. However,
the doping efficiency tends to decrease when the Mn content
increases, either with or without additional Zn flux during the
growth, so that $p\le 10^{19}$~cm$^{-3}$ for $x \ge 0.05$. All
structures with Zn$_{1-x}$Mn$_{x}$Te epilayers exhibited a
temperature-dependent paramagnetic component below 70-80~K. Above
100~K, the signal is temperature independent, providing an adequate
evaluation of the substrate contribution to the total magnetic
moment. Accordingly, a diamagnetic susceptibility
$-1.3\times10^{-6}$~emu, independent of temperature, was determined
for the ZnTe:N epilayer.

Table I presents the hole concentration $p$ as well as the
effective Mn content $x_{\mbox{\small eff}}$ and Curie-Weiss
temperature $T_{\mbox{\small CW}}$, as determined for the studied
samples according to procedures described in the next Section.

\begin{table}[tbp]
\centering \caption[]{Characteristics of the studied samples of
Zn$_{1-x}$Mn$_{x}$Te:N. Hole concentration $p$ was determined from
Hall resistance, whereas effective Mn content $x_{\mbox{\small
eff}}$ and the Curie-Weiss temperature $T_{\mbox{\small CW}}$ from
magnetic susceptibility at $ 4 \le  T \le 20$~K, according to
procedures described in Sec.~III.} \ \\

\begin{tabular}{ccc}
$p$[cm$^{-3}$] &$x_{\mbox{\small eff}}$& $T_{\mbox{\small
CW}}$[K]\\ \hline
 $1.2\times10^{20}$ &0 &  0   \\
$1.2\times10^{20}$  &0.015 &  1.45   \\ $7\times10^{19}$ &0.005
&--\\ $3\times10^{19}$  &0.025 &  2.3
\\ $1.5\times10^{19}$ &0.027 &  2.4\\
$9\times10^{18}$&0.0315&0.75   \\
 $8\times10^{17}$  &0.0285 & $-0.4$
\end{tabular}
\end{table}

\section{EXPERIMENTAL RESULTS}
\subsection{Conductance and magnetoresistance}
The binding energy of an effective mass acceptor in ZnTe, evaluated
from the Baldereschi-Lipari model\cite{Bald73} with the published
values of the Luttinger parameters for ZnTe,\cite{Wagn92} is
59~meV. The same calculation gives the effective Bohr radius,
$a^*\approx 1.3$~nm, which leads to the Mott critical density
$N_{\mbox{\small c}}=(0.26/a^*)^3 =0.8\times10^{19}$~cm$^{-3}$.
Experimentally, substitutional nitrogen forms an even slightly
shallower acceptor, with a binding energy as low as
53~meV.\cite{Grun96} A ZnTe sample with
$p=1.2\times10^{20}$~cm$^{-3}$ is clearly metallic, according to
measurements down to pumped liquid helium temperature [Fig.~1(a)].
The conductivity $\sigma$ of a Zn$_{1-x}$Mn$_{x}$Te sample with $x
= 0.019$ and with the same hole density
($p=1.2\times10^{20}$~cm$^{-3}$) is remarkably identical, if
measured in a magnetic field, as shown in Fig.~1(b), where data
taken in 110~kOe are summarized. The same plot illustrates how the
conductivity decreases when the hole concentration diminishes, and
demonstrates that the temperature dependence of the conductivity
remains weak even for the less doped sample (with
$p=3\times10^{19}$~cm$^{-3}$). In particular, the conductivity of
all samples in the magnetic field stay higher or of the order of
the Mott minimum metallic conductivity $\sigma_{min}\approx
0.03N_{\mbox{\small c}}^{1/3}e^2/\hbar\approx15$ Scm$^{-1}$.

\begin{figure}
\includegraphics*[width=85mm]{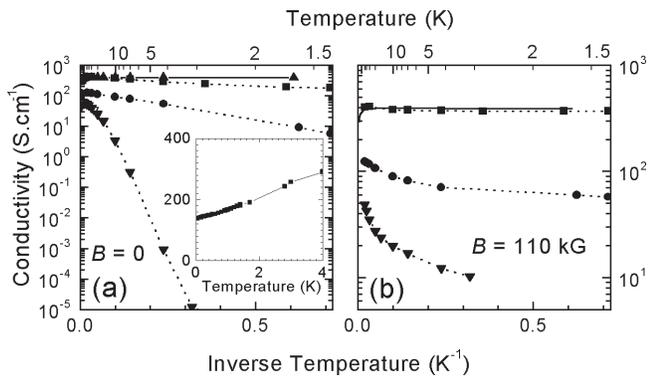}
 \caption[]{Conductivity of ZnTe:N and Zn$_{1-x}$Mn$_{x}$Te:N as a
function of the inverse temperature without (a) and with (b) an
applied magnetic field of 110~kGs; up-triangles stand for ZnTe with
$p=1.2\times10^{20}$~cm$^{-3}$; other symbols are for Zn$_{1-
x}$Mn$_{x}$Te with $x=0.019$  and $p=1.2\times10^{20}$~cm$^{-3}$
(squares); $x=0.005$ and $p=7\times10^{19}$ cm$^{-3}$ (circles),
and $x=0.038$ and $p=3\times10^{19}$~cm$^{-3}$ (down triangles).
The inset to (a) shows the temperature dependence of conductivity
for $x=0.019$ and $p=1.2\times10^{20}$~cm$^{-3}$ in zero magnetic
field. The lines are drawn through experimental points; the solid
line, drawn in both (a) and (b), represents the temperature
dependence of resistivity in the metallic ZnTe sample in zero
magnetic field. The data point to temperature dependent
localization by magnetic disorder, an effect suppressed by a high
magnetic field.}
 \label{fig:rhovst_1}
\end{figure}

The situation is very different without the applied field
[Fig.~1(a)]. The sample with the highest doping level exhibits a
small decrease of conductivity--by a factor of 2.5 between 10~K and
0.1~K, as shown in the inset to Fig.~1(a). Thus, this sample has to
be considered as metallic, also in the absence of the magnetic
field, as its conductivity remains an order of magnitude greater
than $\sigma_{min}$, even at 100~mK. In contrast, the conductivity
of samples with slightly smaller hole concentrations decreases
rather dramatically when lowering the temperature indicating that
the holes become localized at low temperatures in the absence of
the field. Such localization induced by magnetic disorder was also
detected in Bridgman-grown Zn$_{1-x}$Mn$_{x}$Te:P with
$p<10^{19}$~cm$^{-3}$.\cite{Ferr00b} We have checked that no
reentrant metallic behavior occurs in such samples down to 100~mK.
In contrast, a comparison of Figs.~1(a) and 1(b) points to the
presence of a field-induced insulator-to-metal transition, and
indeed a colossal (several orders of magnitude) negative
magnetoresistance is observed at low temperature. Additionally, a
weak positive magnetoresistance in the low field range is also
visible, as reported elsewhere.\cite{Ferr00c}

The above effects, the temperature dependent localization as well
as the positive and negative magnetoresistances, are qualitatively
similar to those observed previously for n-type
DMS.\cite{Furd88,Sawi86,Shap86} In particular, the positive
magnetoresistance results from the giant Zeeman splitting of band
states in DMS, which modifies quantum corrections to the
conductivity at the localization boundary.\cite{Sawi86} Under the
same conditions, a strong spin dependent scattering of itinerant
carriers by ferromagnetic spin puddles (bound magnetic polarons --
BMP) is thought\cite{Sawi86,Glod94} to account for the shifting of
the metal-insulator transition (MIT) towards higher impurity
concentrations as well as for the rapid increase of the resistivity
when decreasing the temperature, and the associated unusually
strong negative magnetoresistance. The latter is enhanced by the
increase of the carrier kinetic energy, which results from their
redistribution between the spin subbands, an effect particularly
important in p-type DMS.\cite{Furd88,Wojt86}

In conclusion, the studies of conductance as a function of
temperature and magnetic field reveal qualitatively similar
properties of n-type and p-type II-VI DMS in the vicinity of the
MIT. In particular, the temperature dependent localization and
negative magnetoresistance indicate that the efficiency of spin
disorder scattering increases at low temperature. This means that
ferromagnetic correlation grows when the temperature decreases.
However, the correlation length of some ferromagnetic puddles has
to remain small, of the order of the de Broglie wavelength of the
itinerant holes, to result in efficient scattering of the hole
spins.

\subsection{ Hall resistivity}

Figure 2 shows the Hall resistivity $\rho_{xy}$ measured at various
temperatures for the highly doped Zn$_{0.981}$Mn$_{0.019}$Te
sample. The quoted hole concentration is deduced from the slope of
the room temperature Hall resistance. We found that $\rho_{xy}$ is
linear in the magnetic field and temperature independent down to
150~K. In the case of the ZnTe sample, this normal Hall effect
$\rho_{xy}$ linear in the field $H$ and temperature independent, is
observed down to 1.6~K (not shown). By contrast, in the case of
Zn$_{1-x}$Mn$_{x}$Te, when decreasing the temperature below 100~K,
one observes first an increase of the slope of the Hall resistance,
and then a strong non-linearity. This "extraordinary" (or
"anomalous") spin-dependent Hall effect $\rho_{xy}^{(an)}$ has
already been discussed in Ref.~\cite{Ferr00c}. It is clearly
observed in III-V DMS,\cite{Mats98} but not in n-doped II-VI
DMS.\cite{Shap86} Its large magnitude stems from the importance of
the spin-orbit coupling in the valence band and from the large
polarization of the hole liquid. The latter results from the giant
Zeeman splitting of the four hole subbands, which is proportional
to the magnetization of the Mn spins. At low temperature and high
field, the Mn or the hole spin polarization saturates, and then the
Hall resistivity exhibits again a linear dependence on the applied
field, with the same slope as at room temperature. Thus, while the
spin-dependent component is too large to allow us to determine the
hole density at low temperatures and in small fields, its magnitude
becomes negligibly small at room temperature, or at low-temperature
in high fields. For these two cases, the slope of the Hall
resistance was found to be identical, giving unambiguously the
value of the hole density.

\begin{figure}
\includegraphics*[width=80mm]{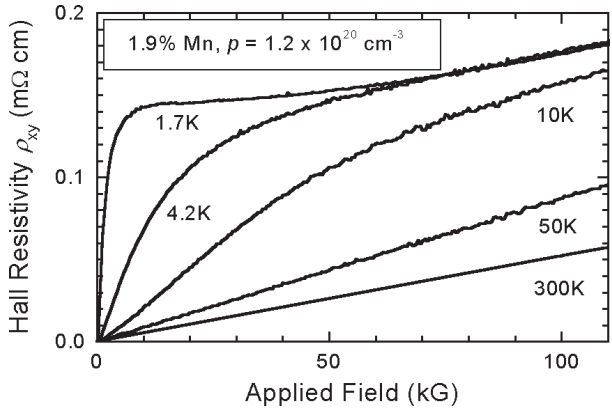}
 \caption[]{Hall resistivity versus magnetic field at different
temperatures, from room temperature down to 1.7~K in metallic
p-Zn$_{0.981}$Mn$_{0.019}$Te:N . The nonlinear temperature
dependent component is assigned to the extraordinary Hall effect,
which strongly increases on approaching the ferromagnetic phase
transition (see Fig.~4).} \label{fig:rhoxy_2}
\end{figure}

In the case of less doped samples, we could measure the Hall
resistivity down to typically 10~K, with the same conclusions, {\it
i.e.}, (i) the normal Hall effect dominates at temperatures above
150~K; (ii) the Hall resistivity varies linearly with the magnetic
field at low temperature in sufficiently large magnetic fields, and
(iii) a strong spin-dependent component appears at weak magnetic
fields and at low temperature, though its accurate determination in
this region is hampered by the large value of the resistance and a
strong magnetoresistance. As mentioned above, the Hall resistance
provides direct information on the degree of spin polarization
$\cal{P}$ of the carrier liquid. However, depending on the dominant
mechanism leading to the extraordinary Hall effect,\cite{Lero72}
$\cal{P}$ is proportional either to the Hall angle, ${\cal{P}} \sim
(\rho_{xy}-\rho_{xy}^{(o)})/ \rho_{xx}$ ("skew scattering"
mechanism) or to the off-diagonal conductivity component,
${\cal{P}} \sim (\rho_{xy} - \rho_{xy}^{(o)})/ \rho_{xx}^2$ ("side
jump" effect). Accordingly, the form of  ${\cal{P}}(T,B)$ deduced
from the Hall and resistivity measurements depends on the assumed
model. The work aiming in elucidating the actual origin of the
extraordinary Hall effect in p-type DMS is under way.

\subsection{Magnetic properties}
\subsubsection{Antiferromagnetic superexchange}

The magnetic properties of undoped Zn$_{1-x}$Mn$_{x}$Te layers and
bulk crystals are well known and the magnetization $M$ in the
magnetic field $H$ can be described by a modified Brillouin
function B$_S$,\cite{Furd88,Gaj79}
\begin{equation}
M=Sg\mu_{\mbox{\small B}}N_0x_{\mbox{\small
eff}}\mbox{B}_S\left[\frac{Sg\mu_{\mbox{\small
B}}H}{k_{\mbox{\small B}}(T+T_{\mbox{\small AF}})}\right].
\end{equation}
The corresponding low-field susceptibility is given by
\begin{eqnarray}
\label{eqn:chin}
 \chi_{\mbox{\small Mn}}&=&C_0x_{\mbox{\small eff}}/(T+T_{\mbox{\small AF}}),
\\ C_0&=&S(S+1)g_{\mbox{\small Mn}}^2\mu_{\mbox{\small B}}^2N_0/3k_{\mbox{\small B}},
\end{eqnarray}
where the Mn spin $S = 5/2$ and the Land\'e factor $g_{\mbox{\small
Mn}}=2.0$; the density of cation sites in ZnTe is
$N_0=1.76\times10^{22}$~cm$^{-3}$. The two empirical parameters
$x_{\mbox{\small eff}} < x$ and $T_{\mbox{\small AF}}> 0$ take into
account the antiferromagnetic superexchange interaction between the
Mn spins. At temperatures $T< 20$~K, the effective density of spin
$x_{\mbox{\small eff}}N_0$ is smaller than the density of Mn ions
$xN_0$  since the nearest-neighbor (n.-n.) Mn pairs are blocked
antiparallel due to their strong superexchange interaction
$J_1/k_{\mbox{\small B}}$.\cite{Furd88} Only "free spins", which
are not involved in these n.-n. pairs, contribute to
$x_{\mbox{\small eff}}$. The phenomenological Curie-Weiss
temperature $T_{\mbox{\small AF}}$ describes the effect of
antiferromagnetic interactions between more distant Mn pairs. This
approach is valid over a field and temperature range, which is
relevant for the present study. A fitting to a set of data from the
literature\cite{Bari85} leads us to (Fig.~3):
\begin{eqnarray}
\label{eqn:eff} x_{\mbox{\small
eff}}&=&x(0.26\mbox{e}^{-43.3x}+0.73\mbox{e}^{-6.2x} +0.01),\\
\label{eqn:taf} T_{\mbox{\small AF}}\mbox{[K]}&=&58x-150x^2.
\end{eqnarray}
The variation of $x_{\mbox{\small eff}}$ is the same as in
Cd$_{1-x}$Mn$_{x}$Te,\cite{Gaj94} which comes as no surprise since
a statistical evaluation of concentration of unpaired spins
explains satisfactorily the experimental
findings.\cite{Shap84,Fata94}

\begin{figure}
\includegraphics*[width=85mm]{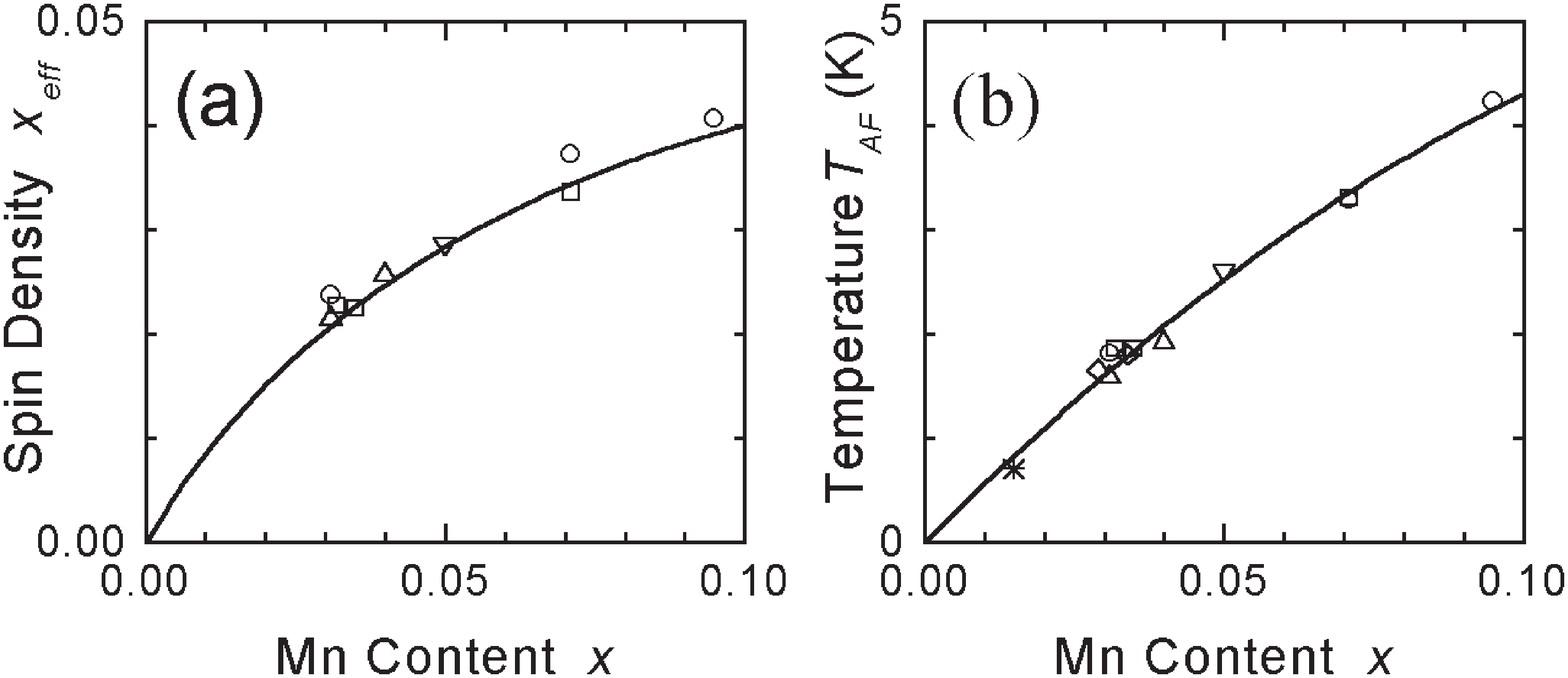}
 \caption[]{Empirical values of parameters characterizing
antiferromagnetic interactions in undoped Zn$_{1-x}$Mn$_{x}$Te: (a)
effective content $x_{\mbox{\small eff}}$ of unpaired Mn spins and
(b) antiferromagnetic Curie-Weiss temperature $T_{\mbox{\small
AF}}$ determined by interactions between non-nearest-neighbors
spins. Lines present fits of Eqs.~(\ref{eqn:eff}) and
(\ref{eqn:taf}) to experimental data (symbols) taken from Ref.~31
(squares: Barilero {\it et al.}; down-triangles: Lascaray {\it et
al.}; up-triangles: Shapira {\it et al.}; circles: Twardowski {\it
et al.}; star: present study, magnetooptical spectroscopy for an
undoped sample).} \label{fig:xefftaf_3}
\end{figure}

\subsubsection{ Metallic regime --- ferromagnetic phase transition}

Experimental results demonstrating the presence of the hole-induced
ferromagnetism in the metallic Zn$_{0.981}$Mn$_{0.019}$Te:N sample
are summarized in Figs.~4 and 5. The susceptibility is determined
from the value of magnetization at 1~kOe above 5~K, and from the
Arrott plots ($M^2$ vs. $H/M$ plots) at lower temperatures, for
which the curvature of the Brillouin function becomes significant.
The inverse magnetic susceptibility plotted as a function of
temperature (Fig.~4) gives the value of the Curie constant $C_0$,
which corresponds to $x_{\mbox{\small eff}}  = 0.015$. According to
Eq.~(\ref{eqn:eff}) this leads to $x = 0.019$. The data collected
in Fig.~3(b) imply that for such a Mn concentration the
antiferromagnetic superexchange results in the Curie-Weiss
temperature $T_{\mbox{\small CW}} = -1.0$~K, which would be
observed for undoped samples. In contrast, the susceptibility and
Hall resistance data for the metallic Zn$_{0.981}$Mn$_{0.019}$Te:N
sample as depicted in Fig.~4, point to a positive value of the
Curie-Weiss temperature $T_{\mbox{\small CW}} = 1.45 \pm 0.1$~K.
This demonstrates that the itinerant holes mediate ferromagnetic
interactions among the Mn spins, which overcompensate the
antiferromagnetic superexchange. However, the small value of the
resulting Curie-Weiss temperature indicates that the ferromagnetic
and antiferromagnetic interactions are of similar magnitudes. The
competition between them constitutes an important ingredient of the
carrier-induced ferromagnetism in II-VI semiconductors.

\begin{figure}
\includegraphics*[width=80mm]{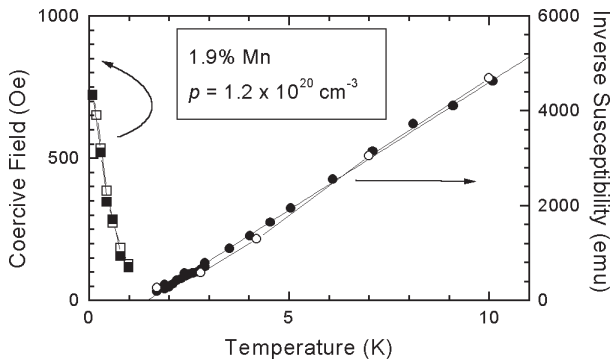}
 \caption[]{Inverse magnetic susceptibility versus temperature for
metallic p-Zn$_{0.981}$Mn$_{0.019}$Te (full circles). Solid line
shows the linear fit, which serves us to determine the effective Mn
content $x_{\mbox{\small eff}}$ and the Curie-Weiss temperature
$T_{\mbox{\small CW}}$ displayed in Table~I. The inverse Hall
resistivity (open circles) and the half-width of hysteresis loops
as determined from Hall resistivity (full squares) and longitudinal
resistivity (open squares) are shown for the same sample. These
data point to a ferromagnetic phase transition in the vicinity of
1.4~K.} \label{fig:chimet_4}
\end{figure}

\begin{figure}
\includegraphics*[width=85mm]{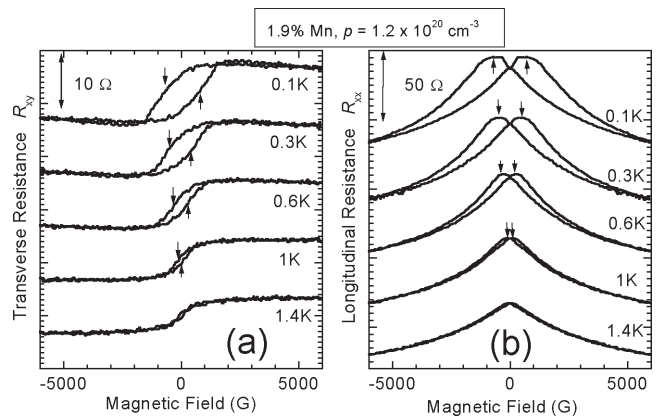}
\caption[]{Low temperature magnetoresistance (a) and Hall
resistance (b) for metallic p-Zn$_{0.981}$Mn$_{0.019}$Te. The
vertical lines mark the width of the hysteresis loops, which is
depicted as a function of temperature in Fig.~4.}
 \label{fig:rhomet_5}
\end{figure}

In order to probe the magnetic ordering below $T_{\mbox{\small
CW}}$, the measurements of the diagonal and Hall resistivities have
been extended down to 100~mK. As shown in Fig.~5, clearly visible
hystereses develop in both $\rho_{xx}$ and $\rho_{xy}$ on lowering
the temperature. This indicates that the low temperature phase is
ferromagnetic. At the same time, the temperature dependence of
hysteresis widths presented in Fig.~4, points to the Curie
temperature $T_{\mbox{\small C}}$ that is only slightly lower than
$T_{\mbox{\small CW}}$. The latter means that the mean-field
approximation constitutes a good starting point for the description
of the ferromagnetism in the studied system, an expected finding in
view of the long-range character of the carrier-mediated exchange
interaction.\cite{Fish72}

\subsubsection{Ferromagnetic interactions mediated by localized holes}

Since the doping efficiency decreases and the magnetic disorder
increases with the Mn concentration, the holes are localized in our
samples with $x > 0.03$.  Actually, the conductance at  $T < 4$~K
drops by several orders of magnitude on crossing the MIT. For
instance, according to Fig.~1, zero-field conductance of
Zn$_{0.962}$Mn$_{0.038}$Te:N with $p=3\times10^{19}$~cm$^{-3}$  is
smaller by more than five orders of magnitude than that of
Zn$_{0.981}$Mn$_{0.019}$Te:N with $p=1.2\times10^{20}$~cm$^{-3}$.
Thus, an interesting question arises how this dramatic difference
in transport properties will affect the hole-mediated ferromagnetic
interactions.

Figure 6(a) shows the temperature dependence of the inverse
magnetic susceptibility of Zn$_{0.962}$Mn$_{0.038}$Te:N with
$p=3\times10^{19}$~cm$^{-3}$. Again the susceptibility was
determined from the magnetization at 1~kOe at higher temperatures
and from the Arrott plots at lower temperatures. A clear
doping-induced positive shift of the Curie-Weiss temperature
$T_{\mbox{\small CW}}$ is put into evidence: we observe
$T_{\mbox{\small CW}}=2.3$~K for a sample with $x=0.038$
($x_{\mbox{\small eff}}=0.025$) and $p=3\times10^{19}$~cm$^{-3}$,
instead of $T_{\mbox{\small CW}}=-T_{\mbox{\small AF}}=-2$~K,
which---according to Fig.~3(b)---is expected for an undoped sample
with the same Mn content.

\begin{figure}
\includegraphics*[width=85mm]{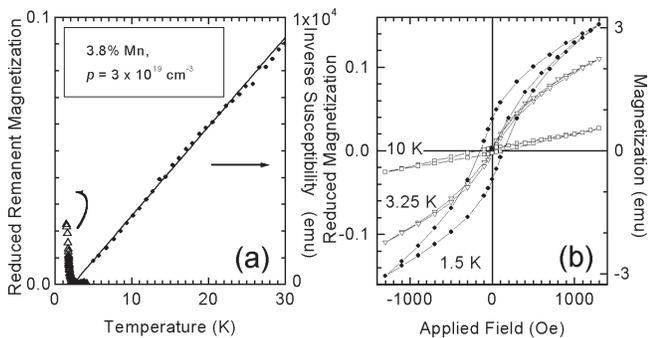}
 \caption[]{(a) Inverse magnetic susceptibility (circles) and remanent
magnetization (triangles) after magnetization cycling up to 1.3~kOe
for p-Zn$_{0.962}$Mn$_{0.038}$Te on the insulator side of the
metal-to-insulator transition. Solid line shows the linear fit,
which serves us to determine the effective Mn content
$x_{\mbox{\small eff}}$ and the Curie-Weiss temperature
$T_{\mbox{\small CW}}$ summarized in Table~I. (b) Magnetization
cycles at various temperatures displayed in the units of the
saturation value.} \label{fig:chimag_6}
\end{figure}

Figure 6(b) shows magnetization loops measured on the sample in
question. A paramagnetic behavior is observed above
$T_{\mbox{\small CW}}$, with a linear dependence of the
magnetization on the magnetic field at high temperature ({\it
e.g.}, at 10~K), and an onset of the field-induced saturation of
magnetization at lower temperature ({\it e.g.}, at 3.2~K).
Magnetization cycles measured at still lower temperature ({\it
e.g.}, at 1.5~K) exhibit hysteresis: The remanent magnetization is
displayed in Fig.~6(a). A slowly decaying component (20\% of the
total signal at the lowest temperature) was observed after shutting
down the field: The values of remanent magnetization plotted in
Fig.~6(a) were measured after 15 min, which corresponded to a
complete decay of the slowly varying component. We may note that
the magnitude of remanent magnetization remains rather small
compared to the total amount of Mn spins deduced from the
susceptibility in the paramagnetic phase. This is only partly due
to the fact that the applied field remains low and these are
probably only minor cycles. Such a small magnitude may indicate
also that the easy axis is out of the plane. However, effects due
to magnetic disorder (competition of ferromagnetic and
antiferromagnetic interactions) as well as due to electrostatic
disorder (mesoscopic fluctuations in the density-of-states) might
also be present and drive the system to a spin-glass phase or to a
ferromagnetic phase with atypically long relaxation
times.\cite{Wolf96}

Figure 7 shows the temperature dependence of the inverse magnetic
susceptibility for two Zn$_{1-x}$Mn$_{x}$Te layers with an
approximate Mn content $x \approx 0.04$ and even smaller hole
concentrations than those in the samples discussed above. Again,
doping-induced positive shift of the Curie-Weiss temperature
$T_{\mbox{\small CW}}$ is clearly visible, as instead of
$T_{\mbox{\small CW}}=-T_{\mbox{\small AF}}=-2.3$~K expected for $x
= 0.04$, the observed values of $T_{\mbox{\small CW}}$ are 2.4 and
$-0.4$~K for $p=1.5\times10^{19}$~cm$^{-3}$  and
$p=7\times10^{17}$~cm$^{-3}$, respectively.

\begin{figure}
\includegraphics*[width=80mm]{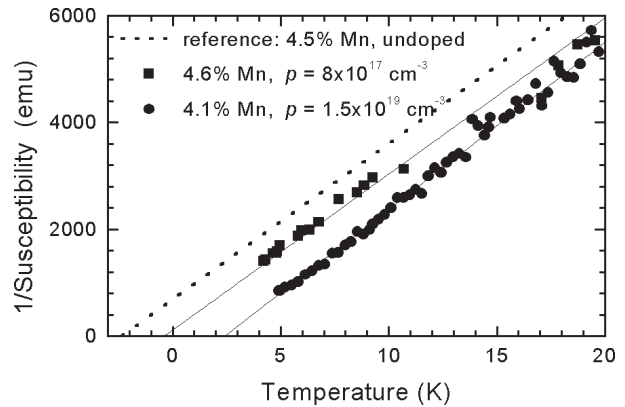}
 \caption[]{Inverse magnetic susceptibility (squares) for two
p-Zn$_{1-x}$Mn$_{x}$Te samples with similar Mn content $x\approx
0.045$ but different hole concentrations, both on the insulator
side of the metal-to-insulator transition. Solid lines show linear
fit, which serves us to determine the effective Mn content
$x_{\mbox{\small eff}}$ and the Curie-Weiss temperature
$T_{\mbox{\small CW}}$ displayed in Table I. The dotted line
presents the dependence expected for an undoped sample with a
similar Mn content.}
 \label{fig:chiins_7}
\end{figure}

We conclude that the Curie-Weiss temperature does not exhibit any
critical behavior on crossing the MIT. This demonstrates the
existence of ferromagnetic interactions of a similar magnitude on
both sides of the MIT. At the same time, it is possible that a
state with an exotic space or time spin correlation, more complex
than that of a simple collinear ferromagnet, develops in the
insulator phase.

\section{MODEL AND MATERIAL PARAMETERS}
\subsection{Zener model and spin susceptibility of carriers}

As discussed above, our study provides the values of Curie-Weiss
$T_{\mbox{\small CW}}$ as a function of the effective spin density
$x_{\mbox{\small eff}}$, deduced from the Curie constant, and the
hole density $p$, determined from the Hall effect measurements.
Knowing $x_{\mbox{\small eff}}$, we determine form
Eqs.~(\ref{eqn:eff}) and (\ref{eqn:taf}) the actual Mn content $x$
and $T_{\mbox{\small AF}}$. The latter would be observed for
undoped samples with the same composition $x$, and characterizes
the strength of antiferromagnetic interactions between Mn spins
more distant than the nearest neighbor pairs. In DMS, the coupling
between the hole of spin $\bm{s}$ and position $\bm{r}$, and the Mn
spin $\bm{S}_i$ localized at $\bm{R}_i$, is well described by a
local exchange interaction\cite{Furd88,Gaj79,Gaj78}
\begin{equation}
H_{p-d}=-\beta\bm{s}\cdot\bm{S}_i\delta(\bm{r}-\bm{R}_i ),
 \end{equation}
where the spin-hole exchange energy is known from
magneto-spectroscopy to be $\beta N_0 = -1.1$~eV in
Zn$_{1-x}$Mn$_{x}$Te.\cite{Twar84} This interaction can be used to
evaluate the energy $J_{ij}$ of the carrier-mediated RKKY exchange
coupling between the Mn spins $i$ and $j$. This coupling exhibits
the well-known oscillations as a function of the distance between
the Mn pairs with the period determined by the Fermi wavevector
$k_{\mbox{\small F}}^{-1}$. Then, in order to obtain the mean-field
value of $T_{\mbox{\small CW}}$, the interaction energy has to be
averaged over the distribution of the Mn spins. This will be
elaborated later. Now we consider the limit of low carrier density,
in which the mean Mn-Mn distance is small with respect to
$k_{\mbox{\small F}}^{-1}$. In such a case, the continuous-medium
limit of the RKKY model, {\it i.e.}, the Zener model,\cite{Zene50}
can be used.\cite{Diet97,Diet00a,Cibe99} The mean-field version of
this model can be summarized as follows.

First we use the molecular-field and virtual-crystal approximations
to relate the spin polarization of the hole liquid to the Mn
magnetization $\bm{M}$. The result is
\begin{eqnarray}
\label{eqn:sz}
 \nonumber \langle s_z \rangle &=&
\beta\tilde{\chi}_{\mbox{\small h}}\sum_i\langle
S_z^{i}\rangle\delta(\bm{r} - \bm{R}_i)\\
 &=&\beta\tilde{\chi}_{\mbox{\small h}}x_{\mbox{\small eff}}N_0\langle S_z\rangle,
\end{eqnarray}
where the first-line summation runs over a discrete distribution of
the Mn spins, and is replaced in the second line by an average spin
projection on the direction of $\bm{M}$, taken here as $z$-axis.
The quantity $\tilde{\chi}_{\mbox{\small h}}$ is the carrier spin
susceptibility, which in the absence of the spin-orbit interaction
would correspond to the Pauli magnetic susceptibility with the
$g\mu_{\mbox{\small B}}$ factor omitted.

We also adopt the mean-field approximation (MFA) to obtain $M$
\begin{eqnarray}
\label{eqn:mfa}
 \nonumber M/g_{\mbox{\small Mn}}\mu_{\mbox{\small B}} &=& -x_{\mbox{\small eff}}N_0\langle
S_z\rangle\\
  &=& \tilde{\chi}_{\mbox{\small Mn}}(g_{\mbox{\small Mn}}\mu_{\mbox{\small B}} H  - \beta\langle s_z\rangle),
\end{eqnarray}
where
\begin{equation}
 \tilde{\chi}_{\mbox{\small Mn}}=\frac{\tilde{C}_0x_{\mbox{\small eff}}}{T+T_{\mbox{\small AF}}}\,\,
 {\mbox{and}}\,\,\tilde{C}_0=\frac{S(S+1)N_0}{3k_{\mbox{\small B}}}.
\end{equation}
Combining Eqs.~(\ref{eqn:sz}) and (\ref{eqn:mfa}) we are led
to\cite{Diet97}
\begin{equation}
M=\frac{\chi_{\mbox{\small Mn}}}{1-\tilde{\chi}_{\mbox{\small
Mn}}\tilde{\chi}_{\mbox{\small h}}\beta^2}H=
\frac{C_0x_{\mbox{\small eff}}}{T-T_{\mbox{\small CW}}}H,
\end{equation}
with
\begin{equation}
 T_{\mbox{\small CW}} =
T_{\mbox{\small F}}-T_{\mbox{\small AF}}\,\, \mbox{and}\,\,
T_{\mbox{\small F}}=x_{\mbox{\small
eff}}\tilde{C}_0\beta^2\tilde{\chi}_{\mbox{\small h}},
\label{eqn:tf}
\end{equation}
where, within the MFA, $T_{\mbox{\small CW}}$ is equal to the Curie
temperature $T_{\mbox{\small C}}$. Usually, however, the MFA, which
can also be regarded as a high temperature expansion, gives a
better estimate of $T_{\mbox{\small CW}}$ than of  $T_{\mbox{\small
C}}$.

The above reasoning can easily be generalized to the case of a
phase transition to a spatially-modulated ground-state,
characterized by non-zero magnetization $\bm{M}(\bm{q})$. The
corresponding mean-field value of the ordering temperature
$T_{\mbox{\small C}}(\bm{q})$ is given by the solution of the
equation:\cite{Diet99}
\begin{equation}
\tilde{\chi}_{\mbox{\small Mn}}(\bm{q},T)\tilde{\chi}_{\mbox{\small
h}}(\bm{q},T)\beta^2=1.
\end{equation}

In the case of the conduction band, the periodic part of the Bloch
wave function is S-like, so that spin dynamics is not perturbed by
the spin-orbit interaction. The spin susceptibility of a degenerate
carrier liquid at $q = 0$ is then directly related to the
density-of-states at the Fermi energy,
\begin{equation}
\tilde{\chi}_e=\frac{1}{4}A_{\mbox{\small F}}\rho(E_{\mbox{\small
F}}).
\end{equation}
Here, the prefactor $A_{\mbox{\small F}}$ is the Fermi liquid
parameter that takes into account the enhancement of the spin
susceptibility by the carrier-carrier exchange interactions.
However, in the case of the holes involving P-like states, the
influence of the spin-orbit coupling has to be taken into
account.\cite{Diet97,Diet00a}

\subsection{Spin susceptibility of holes}

The valence band in ZnTe is characterized by a rather strong
spin-orbit coupling, $\Delta_{\mbox{\small o}}=
0.96$~eV.\cite{Herm84} Hence at the center of the Brillouin zone
the $\Gamma_7$ split-off band is well separated from the $\Gamma_8$
quadruplet, and for small wavevectors, the dispersion can be
calculated by using the $4\times4$ Luttinger Hamiltonian. We adopt
the experimental values of the Luttinger parameters\cite{Wagn92}
$\gamma_1=3.8$, $\gamma_2=0.72$, and $\gamma_3=1.3$, and use the
spherical approximation,\cite{Bald73} {\it i.e.}, we replace
$\gamma_2$ and $\gamma_3$ by
$\gamma_s=\frac{2}{5}\gamma_2+\frac{3}{5}\gamma_3$. In this
approximation, the values of the heavy-hole mass $m_{\mbox{\small
hh}}=0.60m_0$ and the light-hole mass $m_{\mbox{\small
lh}}=0.17m_0$, are independent of the direction of the wave vector
$\bm{k}$.

We are interested in the response function of the holes for the
molecular field applied along the $z$ direction. However, at $k\ne
0$, the periodic parts of the Bloch functions are the eigenstates
of $\bm{J}\cdot\bm{k}/k$, i.e, the quantization axis is along
$\bm{k}$. The corresponding eigenfunctions, $|\pm
3/2,\bm{k}\rangle$ for the heavy holes and $|\pm 1/2,\bm{k}\rangle$
for the light holes, can be obtained from the eigenfunctions of
$J_z$, $|\pm 3/2\rangle_z$ and $|\pm 1/2\rangle_z$, by applying the
corresponding rotation matrix within the $\Gamma_8$ quadruplet,
identical to the rotation matrix within a $J=3/2$ quadruplet. The
important point here is that $s_z$ in the basis of the
eigenfunctions $|\pm 3/2,\bm{k}\rangle$ and $|\pm
1/2,\bm{k}\rangle$  reads
\begin{equation}
s_{z} = \left[\begin{array}{cccc}
\frac{\cos\Theta}{2}&0&-\frac{\sin\Theta}{2\sqrt{3}}&0\\
0&-\frac{\cos\Theta}{2}&0&-\frac{\sin\Theta}{2\sqrt{3}}\\
-\frac{\sin\Theta}{2\sqrt{3}}&0&\frac{\cos\Theta}{6}&-\frac{\sin\Theta}{3}\\
0&-\frac{\sin\Theta}{2\sqrt{3}}&-\frac{\sin\Theta}{3}&-\frac{\cos\Theta}{6}
\end{array} \right],
\end{equation}
where $\Theta$ is the polar angle of the wavevector $\bm{k}$. There
are no matrix elements between states with different $\bm{k}$
vectors.

The longitudinal component of the hole spin susceptibility is
\begin{equation}
\tilde{\chi}_{\mbox{\small
h}}(\bm{q})=2\sum_{i,j,\bm{k}}\frac{\left|\langle
i,\bm{k}|s_z|j,\bm{k}+\bm{q}\rangle\right|^2}{E_{j,\bm{k}+\bm{q}}-
E_{i,\bm{k}}}f(E_{i,\bm{k}})\left[1-f(E_{j,\bm{k}+\bm{q}})\right],
\label{eqn:chiq}
\end{equation}
where $|i,\bm{k}\rangle$ are the periodic part of the Bloch
functions, $E_{i,\bm{k}}=\hbar^2k^2/2m_{\mbox{\small hh}}$ or
$\hbar^2k^2/2m_{\mbox{\small lh}}$, and $f(E_{i,\bm{k}})$ is the
Fermi-Dirac distribution function for the corresponding hole
subbands. The RKKY interaction energy is proportional to the
Fourier transform of Eq.~(\ref{eqn:chiq}). In the framework of the
Zener model of ferromagnetism, we are interested in
$\tilde{\chi}_{\mbox{\small h}}(\bm{q}=0)$. The result that is
obtained after a straightforward calculation for the degenerate
hole gas assumes the form
\begin{equation}
\tilde{\chi}_{\mbox{\small h}}=\frac{1}{4}A_{\mbox{\small
F}}\rho(E_{\mbox{\small F}})\left[\frac{1}{3}
+\frac{8}{9}\frac{m_{\mbox{\small hh}}^{3/2}m_{\mbox{\small
lh}}-m_{\mbox{\small lh}}^{3/2}m_{\mbox{\small
hh}}}{(m_{\mbox{\small hh}}- m_{\mbox{\small lh}})(m_{\mbox{\small
hh}}^{3/2}+m_{\mbox{\small lh}}^{3/2})}\right], \label{eqn:chih}
\end{equation}
 where the Fermi liquid parameter $A_{\mbox{\small F}}$, to be discussed later,
 represents the effect of the hole-hole
interaction, and
\begin{equation}
\rho(E_{\mbox{\small F}})=(m_{\mbox{\small
hh}}^{3/2}+m_{\mbox{\small
lh}}^{3/2})^{2/3}(3\pi^2p)^{1/3}/\pi^2\hbar^2. \label{eqn:dos}
\end{equation}

The two terms in the square brackets in Eq.~(\ref{eqn:chih})
describes two distinct effects of the spin-orbit interaction on the
hole spin susceptibility:\\ - the first arises from terms which are
diagonal within the heavy-hole and within the light-hole subbands.
Thus, the spin-orbit interaction reduces intra-band hole spin
polarization by a factor of three. Another, more intuitive,
description is obtained by noting that applying the molecular field
$B$ along the $z$ direction splits the Fermi surface into two (up
and down) surfaces, with the Zeeman splitting of the heavy holes
proportional to $B\langle\pm 3/2,\bm{k}|s_z|\pm
3/2,\bm{k}\rangle=\pm B\cos\Theta$,\cite{Gaj78} as shown in Fig.~8.
The spin polarization is obtained by integrating $s_z$ over the
states between the two Fermi surfaces, both $s_z$ and the
population depending on $\Theta$. This means, in particular, that
the spin polarization of the holes, {\it i.e.}, the spin component
of their magnetization, is not simply proportional to the
difference in the up and down populations, as it would be the case
for the electrons in the conduction band. By integrating the sum of
the square of the relevant matrix elements over the solid angle we
obtain the reduction factor
$\int_0^{\pi}\cos^2\Theta\mbox{d}\cos\Theta/\int_0^{\pi}\mbox{d}
\cos\Theta=1/3$. Hence, this reduction factor expresses the fact
that the heavy-hole spin is quantized along the direction of
$\bm{k}$, so that exchange splitting vanishes for
$\bm{k}\perp\bm{B}$. A similar procedure for the light-hole subband
leads to the same reduction factor 1/3.\\ - The second term arises
from non-diagonal terms coupling the heavy and light-hole subbands,
that is, it represents interband spin polarization. Then, the Fermi
distribution functions define the range of integration, from the
Fermi wavevector of the light-holes to that of the heavy-holes. The
corresponding contribution to $T_{\mbox{\small CW}}$ can be viewed
as the manifestation of the Bloembergen-Rowland
mechanism\cite{Furd88} of the indirect spin-spin exchange
interaction, allowed here by the spin-orbit coupling.

\begin{figure}
\includegraphics*[width=80mm]{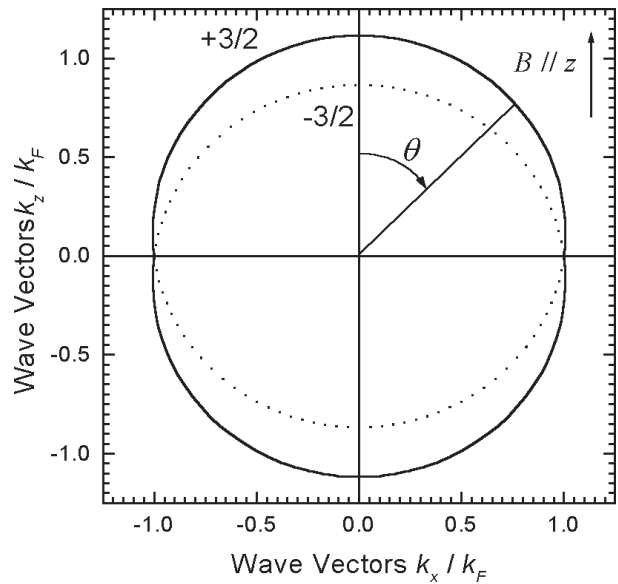}
 \caption[]{Cross section of the heavy hole Fermi surface in presence
of the exchange field (magnetization) applied along z direction.
The plot is shown for $\Sigma_z(B)/E_{\mbox{\small F}}=0.5$,  where
$\Sigma_z(B)$ is the spin-splitting at $k=0$, but the calculation
of the spin susceptibility is performed in the limit of vanishing
$\Sigma_z(B)/E_{\mbox{\small F}}$.}
 \label{fig:fermi_8}
\end{figure}

Introducing into Eq.~(\ref{eqn:chih}) the values of the hole
effective masses corresponding to ZnTe we find that the spin-orbit
interaction reduces the hole spin susceptibility by a factor of
2.1,
\begin{equation}
\tilde{\chi}_{\mbox{\small h}}=\frac{1}{4}\rho(E_{\mbox{\small
F}})\frac{1}{2.1}. \label{eqn:chir}
\end{equation}
It has been checked that $T_{\mbox{\small CW}}$ values resulting
from Eqs.~(\ref{eqn:tf}) and (\ref{eqn:chir}) agree with those
determined by numerical minimization of the free energy, which is
obtained by summing up eigenvalues of the corresponding $4\times4$
Luttinger matrix.\cite{Diet00a,Diet00b} Actually, the theoretical
model developed recently,\cite{Diet00a,Diet00b} allows one to
determine the magnetization as a function of temperature and
magnetic field taking into account non-zero anisotropy, finite
spin-orbit splitting, and biaxial strain within the $6\times6$
Luttinger model of the valence band in tetrahedrally coordinated
semiconductors. We shall compare results of the $4\times4$ and
$6\times6$ models {\it vis-\`a-vis} our experimental data on
$T_{\mbox{\small CW}}$, and show that the difference between their
predictions is slight for the parameters of p-Zn$_{1-x}$Mn$_{x}$Te.

\subsection{Zener vs. RKKY model}

We identify two experimentally important situations, for which the
Zener model, as introduced above, ceases to be valid. The first one
corresponds to the case when an average time of carrier tunneling
between Mn pairs $(Vx_{\mbox{\small eff}}^{1/3})^{-1}$ becomes
significantly longer than the inverse exchange energy $|\beta
N_0|^{-1}$. Here $V$ is the width of the carrier band, and its
magnitude, not the Fermi energy as sometimes suggested, constitutes
the relevant energy scale. For long tunneling times, the
molecular-field and virtual-crystal approximations break down, an
effect detected in Cd$_{1-x}$Mn$_{x}$S.\cite{Beno92}
Double-exchange model constitutes the appropriate description of
the carrier-mediated exchange interaction in the limit $V
\rightarrow 0$.

The second case is that of a large carrier concentration, $n >
x_{\mbox{\small eff}}N_0$. In this region, important changes in the
carrier response function occur at the length scale of a mean
distance between the localized spins. Accordingly, the description
of spin magnetization by the continuous-medium approximation, which
constitutes the basis of the Zener model, ceases to be valid.  In
contrast, the RKKY model is a good starting point in this regime,
as it provides the dependence of the interaction energy of
particular spin pairs as a function of their distance. This makes
it possible to evaluate the system energy for a given distribution
of the localized spins.

We note in passing that in the limit when the continuous-medium
approximation is valid, $n \gg x_{\mbox{\small eff}}N_0$, the
mean-field value of the ordering temperature $T(\bm{q})$ deduced
from the Zener and RKKY model are identical, independently of
microscopic spin arrangement. In particular, in both models, the
Curie temperature is determined entirely by matrix elements that
are diagonal in $\bm{k}$, as non-diagonal terms vanish when the
continuous-medium approximation is valid. If this is not the case,
the two models are equivalent only if the spins are randomly
distributed over a continuum.\cite{Diet97}

Since the RKKY model appears as more general, it is tempting to
adopt it for the description of experimental findings.
Unfortunately, however, in the presence of spin-orbit interaction,
the spin-spin Hamiltonian contains non-scalar pseudo-dipole and
Moriya-Dzialoshinskii terms, whose dependence of the pair distance
seems to be described by non-elementary functions. Briefly, the
RKKY model is technically much more cumbersome than the Zener model
for the holes in the $\Gamma_{8}$ band. On the other hand, in the
case of the sample with the highest doping level, the hole
concentration $p=1.2\times10^{20}$~cm$^{-3}$ becomes comparable
with $x_{\mbox{\small eff}}N_{0} = 1.8\times10^{20}$~cm$^{-3}$.
Moreover, we note that the blocking of n.-n. Mn pairs in the
zero-spin state by antiferromagnetic superexchange not only reduces
the effective spin concentration but also makes the spin
distribution to be highly non-random --- no n.-n. Mn pairs are
involved in the ferromagnetic interactions.

In order to evaluate the resulting effect on the Curie-Weiss
temperature we start from the expression for the energy of the RKKY
exchange interaction between two Mn spins at relative position
$\bm{R}$ induced by carriers described by S-like wave functions
(see, {\it e.g.} Ref.~3),
\begin{equation}
J(\bm{R})=\chi_{\mbox{\small h}}(0)\frac{2k_{\mbox{\small
F}}^3}{\pi}\beta^2\frac{\sin(2k_{\mbox{\small F}}R)-
2k_{\mbox{\small F}}R\cos(2k_{\mbox{\small F}}R)}{(2k_{\mbox{\small
F}}R)^4}. \label{eqn:rkky}
\end{equation}
Since the main contribution to $\tilde{\chi}_{\mbox{\small
h}}(\bm{q})$ in Eq.~(\ref{eqn:chiq}) comes from the heavy hole
band, we use the heavy hole wavevector at the Fermi level for
$k_{\mbox{\small F}}$ in Eq.~(\ref{eqn:rkky}). At the same time,
the prefactor in Eq.~(\ref{eqn:rkky}), the hole spin susceptibility
$\tilde{\chi}_{\mbox{\small h}}$ at $q=0$, is calculated from
Eq.~(\ref{eqn:chih}), so that it takes the complex structure of the
$\Gamma_8$ bands into account.

The mean-field value of ferromagnetic temperature $T_{\mbox{\small
F}}$ is then obtained from the first moment of the distribution of
the pair interaction energies $J(R_{ij})$. Thus, we determine
$T_{\mbox{\small F}}$ by a summation of the interaction energies
$J(R)$ between a given spin at $R = 0$, and all other free Mn spins
distributed over the cation sites of the zinc-blende lattice,
\begin{equation}
T_{\mbox{\small F}}=\frac{S(S+1)}{3k_{\mbox{\small
B}}}\sum_{\mbox{{\small fcc sites}}}P(i)J(\bm{R}_i).
\label{eqn:tfrkky}
\end{equation}
Here, the occupation probability is $P(i) = 0$ for the nearest
neighbors to the origin (since the point $R = 0$ is occupied by the
Mn spin, the n.-n. spin would be blocked antiferromagnetically),
and $P(i)=x_{\mbox{\small eff}}$ for all other fcc sites. As
mentioned above, replacing the discrete sum in
Eq.~(\ref{eqn:tfrkky}) by a continuous integration with the uniform
density $x_{\mbox{\small eff}}$ would lead to Eq.~(\ref{eqn:tf}).
We shall compare those two models for $T_{\mbox{\small F}}$ as well
as their ability to describe the experimental data in Sec.~V.
Finally, we note that on increasing the carrier concentration the
second moment of the distribution $J(R_{ij})$ grows faster than the
first moment. The former determines the spin-freezing temperature
$T_{\mbox{\small g}}$, which at some point may become higher than
$T_{\mbox{\small F}}$. We evaluate that $T_{\mbox{\small g}} <
T_{\mbox{\small F}}$ for the samples studied here. Such a
cross-over from the ferromagnetic to spin-glass phase as a function
of $p/xN_0$ has, in fact, been observed in IV-VI DMS.\cite{Egge94}
For even greater values of $p/xN_0$, the Kondo temperature
$T_{\mbox{\small K}}$ may become higher than $T_{\mbox{\small g}}$,
so that screening of the spins by the carriers will occur at  $T <
T_{K}$ if $T_{\mbox{\small K}} > T_{\mbox{\small g}}$.

\subsection{Effects of disorder and carrier-carrier interactions}

The present experimental results address the question of how
disorder in the electronic subsystem affects the carrier-mediated
interaction among the localized spins. In the framework of the RKKY
model, the mean free path $l$ sets an upper distance of Mn pairs
contributing to the first moment of the distribution $J(R_{ij})$.
This leads to a reduction of $T_{\mbox{\small F}}$ by $(1 -
\pi/4k_{\mbox{\small F}}l)$ for $k_{\mbox{\small F}}l \gg
1$.\cite{Diet97} In terms of the Zener model, this reduction factor
represents the scattering broadening of the thermodynamic
density-of-states $\rho(E_{\mbox{\small F}})$.\cite{Diet97,Koss00}
At the same time, the second moment, and thus $T_{\mbox{\small
g}}$, is only affected by spin-dependent scattering.\cite{Zyuz86}

Interestingly enough, neither $l$ nor $\rho(E_{\mbox{\small F}})$
exhibit critical behavior at the MIT -- typically $l \approx
k_{\mbox{\small F}}^{-1}$ at criticality when approaching the MIT
from the metallic side. In contrast, the localization length $\xi$
diverges at the MIT and, according to the scaling picture,
decreases gradually towards the Bohr radius $a^*$ deeply in the
insulator phase. Then, the ferromagnetic exchange is mediated by
carriers thermally excited to the band\cite{Pash79} or by
spin-dependent coupling between BMP.\cite{Wolf96} However, over a
wide range of dopant concentrations, the average value of $\xi$ is
significantly larger than both $a^*$ and $l$. We expect that such
weakly localized carriers carry efficiently the spin-spin coupling.
Since distances smaller than $l$ are important for ferromagnetism,
carrier localization will have a minor influence on the value of
$T_{\mbox{\small F}}$ for such an exchange mechanism.

At the same time, however, large mesoscopic fluctuations in the
local values of $\rho(E_{\mbox{\small F}})$ and $\xi$, are expected
in the vicinity of the MIT. This will introduce additional
randomness in the system, the corresponding correlation length
being of the order of either spin coherence, thermal or
localization length, each presumably much longer than that
characterizing magnetic disorder. In a simplistic two-fluid
picture,\cite{Diet00a,Diet00b} we envisage that mosaics of
ferromagnetic and paramagnetic regions appear at the Curie
temperature. The latter corresponds to regions not visited by the
carriers. The former contain delocalized or weakly localized
carriers.  Such carriers set a long-range ferromagnetic correlation
between the Mn spins, including those contributing to BMP that are
formed around singly occupied impurity-like states. According to
this model, the ferromagnetic portion of the material, and thus the
magnitude of spontaneous magnetization, grows with the dopant
concentration.

Finally, we consider the enhancement of the tendency towards
ferromagnetism by the carrier-carrier interactions. Within the
Zener model, this enhancement is described by the Fermi liquid
parameter $A_{\mbox{\small F}}$. The value $A_{\mbox{\small
F}}\approx 1.2$ was calculated by a local spin-density approach for
a 3D carrier liquid of similar density.\cite{Jung99} We adopt this
value here, as its enhancement by disorder is probably unimportant
due to efficient spin-flip scattering in our system.

\section{COMPARISON OF EXPERIMENTAL AND THEORETICAL RESULTS}

In order to compare experimental and theoretical results for
samples with various Mn content $x$ and hole concentrations $p$, we
introduce a normalized value of the ferromagnetic temperature,
\begin{equation}
\tilde{T}_{\mbox{\small F}}=(T_{\mbox{\small CW}}+T_{\mbox{\small
AF}})/10^2x_{\mbox{\small eff}}, \label{eqn:tfn}
\end{equation}
where $T_{\mbox{\small CW}}$ and $x_{\mbox{\small eff}}$ are
determined from the measurements of the magnetic susceptibility as
a function of temperature, whereas the corresponding values of
$T_{\mbox{\small AF}}$ are calculated from Eq.~(\ref{eqn:taf}). The
actual values of $T_{\mbox{\small CW}}$ and $x_{\mbox{\small eff}}$
for the studied samples are summarized in Table~I (Sec.~II).
According to the Zener model (Eq.~\ref{eqn:tf}), the normalized
ferromagnetic temperature, as defined by Eq.~(\ref{eqn:tfn}), does
not depend on the Mn content.

Figure 9 presents experimental and theoretical values of
$\tilde{T}_{\mbox{\small F}}$ as a function of $p$. The studied
hole concentration range covers both sides of the MIT, the sample
with the highest $p$ values being metallic. Remarkably, the
presence of the MIT appears to have no effect on the experimental
values of $\tilde{T}_{\mbox{\small F}}(p)$. This substantiates the
view that the two phenomena, carrier-mediated ferromagnetic
interactions and carrier localization, are sensitive to carrier
wave functions at different length scales: the former shorter while
the latter longer than the mean free path. This encourages us to
interpret the data disregarding localization effects. Furthermore,
we shall neglect the influence of scattering on the
density-of-states, an approximation that should be relaxed once
appropriate information would be available.

\begin{figure}
\includegraphics*[width=80mm]{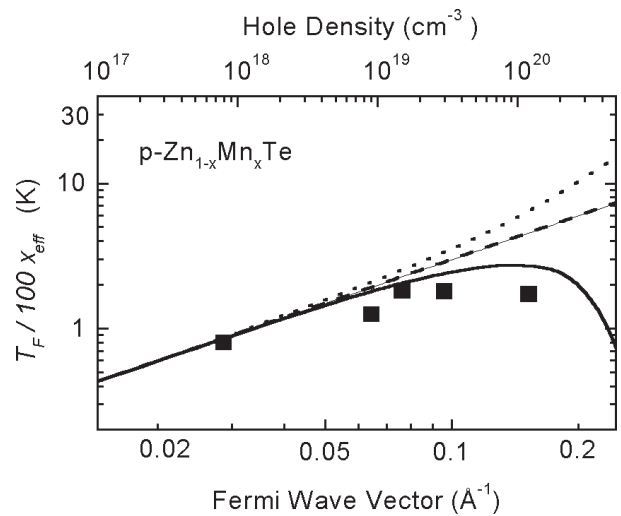}
 \caption[]{Experimental (symbols) and calculated (lines) normalized
ferromagnetic temperature, $T_{\mbox{\small F}}/10^2x_{\mbox{\small
eff}}$, versus the wave vector at the Fermi level. The
corresponding hole densities are indicated in the top scale. Dashed
line: Zener model with the hole dispersion calculated from the
$4\times4$ Luttinger spherical model for the $\Gamma_8$ band;
dotted line: Zener model including the coupling between the
$\Gamma_8$ and $\Gamma_7$ bands ($6\times6$ Luttinger model,
Ref.~14); solid line: the RKKY and $6\times6$ Luttinger model for
$x_{\mbox{\small eff}}=0.015$, taking into account the effect of
the antiferromagnetic interactions on statistical distribution of
unpaired Mn spins.}
 \label{fig:theory_9}
\end{figure}

The dotted line in Fig.~9 represents results of the analytic
calculation within the Zener model, employing $4\times4$ spherical
Luttinger Hamiltonian, Eqs.~(\ref{eqn:tf}) and (\ref{eqn:chih}).
Similarly, the dashed line was obtained numerically from the
$6\times6$ model.\cite{Diet00a,Diet00b} Due to the large magnitude
of spin-orbit splitting in ZnTe, the difference between the models
is slight, even for relatively large hole concentrations. It is
seen that the Zener model describes correctly the ferromagnetic
temperature in the region of low hole concentrations but it
predicts a significantly too large value of
$\tilde{T}_{\mbox{\small F}}$  for the sample with
$p=1.2\times10^{20}$ cm$^{-3}$. As already mentioned, in this
sample $x_{\mbox{\small eff}}$ = 0.015, so that the inverse Fermi
wavevector becomes comparable to the mean Mn-Mn distance,
$d_{Mn-Mn}k_{\mbox{\small F}} = 1.4$. As a result the
continuous-medium approximation, inherent to the Zener model,
ceases to be valid. The solid line in Fig.~9 depicts numerical data
obtained within the RKKY model, Eqs.~(\ref{eqn:rkky}) and
(\ref{eqn:tfrkky}) for $x_{\mbox{\small eff}}$ = 0.015, with the
long-wavelength hole spin susceptibility computed by the $6\times6$
Luttinger model. A significant reduction of
$\tilde{T}_{\mbox{\small F}}$ for large $p$, consistent with the
experimental findings, is clearly visible. This reduction reflects
the important effect of the antiferromagnetic n.-n. Mn pairs on the
distribution of the free spins. In view that theory is developed
with no adjustable parameters, we conclude that the main processes
accounting for the magnitude of the Curie-Weiss temperature are
well understood in p-Zn$_{1-x}$Mn$_{x}$Te.

Figure 10 presents the dependence of magnetization on the magnetic
field for one non-metallic and one metallic sample. The dependence
of magnetization on the magnetic field is in accord with the
expectation of the mean-field theory in the case of the metallic
sample. In contrast, the magnitude of magnetization observed in the
non-metallic sample in the weak fields is about two times smaller
than expected. This substantiates the conjecture about the phase
separation into the regions with a different degree of Mn spin
polarization on the insulator side of the MIT.

\begin{figure}
\includegraphics*[width=85mm]{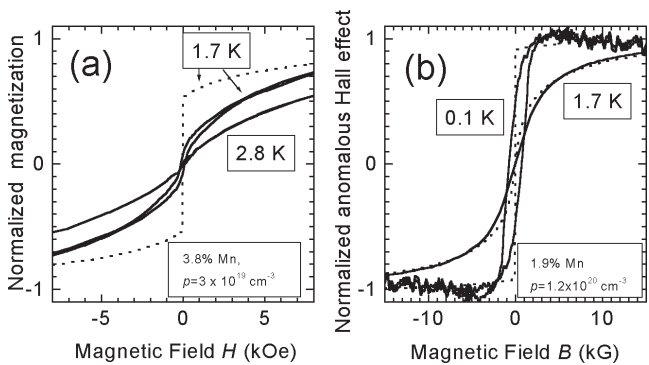}
 \caption[]{ Low-temperature magnetization as a function of the magnetic
field measured directly for non-metallic
p-Zn$_{0.962}$Mn$_{0.038}$Te sample (a) and determined from the
Hall data for metallic p-Zn$_{0.981}$Mn$_{0.019}$Te sample (b),
whose magnetic susceptibilities are shown in Figs.~6 and 4,
respectively. The doted lines present results of the mean-field
calculation, which indicate that only a part of Mn spins contribute
to ferromagnetic order in the insulating sample.}
\label{fig:magvst_10}
\end{figure}

\section{SUMMARY AND OUTLOOK}

The progress in nitrogen doping of Zn$_{1-x}$Mn$_{x}$Te by MBE
achieved in course of this work has made it possible to grow II-VI
DMS, in which hole kinetic energy is high enough to overcompensate
electrostatic and magnetic disorder, so that the metal phase exists
down to the millikelvin temperature range. In such a sample the
doping-induced ferromagnetic ordering has been put into the
evidence. Two factors have been identified, which make the Curie
temperature relatively low, $T_{\mbox{\small C}}\approx 1.5$~K at
$p = 1.2\times10^{20}$ cm$^{-3}$ and $x =0.019$, despite the rather
large values of the effective mass and p-d exchange integral.
First, the spin-orbit interaction in the valence band leads to more
than two-fold reduction of $T_{\mbox{\small C}}$. Second, the
superexchange antiferromagnetic interactions have been found to
lower $T_{\mbox{\small C}}$ even further. In particular, the
residual antiferromagnetic interactions between more distant Mn
pairs contribute to the reduction of $T_{\mbox{\small C}}$.
Moreover, the presence of magnetically inert nearest-neighbor Mn
pairs not only lowers the effective Mn concentration but also makes
the antiferromagnetic portion of the RKKY interaction to become
more significant. The resulting competition between the
ferromagnetic and antiferromagnetic interactions is expected to
grow with $p/x$, and may ultimately result in the transition to a
spin-glass phase. Alternatively, with decreasing $x$ at given $p$,
the Kondo effect may show up.

We note that the present results, as well as those for
Ga$_{1-x}$Mn$_{x}$As,\cite{Diet00a,Diet00b} make it possible to
single out dominant effects accounting for differences between
temperatures of ferromagnetic ordering in particular families of
magnetic semiconductors.\cite{Diet00a,Diet00b} One of the effects
is the magnitude of the spin-orbit splitting $\Delta_{\mbox{\small
o}}$ in the valence band, which---if much greater than the Fermi
energy---reduces $T_{\mbox{\small C}}$. This is the case of
p-Zn$_{1-x}$Mn$_{x}$Te. In contrast, due to the smaller value of
$\Delta_{\mbox{\small o}}$ in Ga$_{1-x}$Mn$_{x}$As, the reduction
ceases to be important. Moreover, the mixing between $\Gamma_8$ and
$\Gamma_7$ bands enlarges the density-of-states at the Fermi level.
Another important aspect of III-V DMS is associated with the fact
that Mn ions supply both spins and holes. This results in a large
Coulomb potential at closely lying Mn pairs. Accordingly, such a
complex binds a hole, which mediates a strong ferromagnetic
coupling that overcompensates the intrinsic antiferromagnetic
interaction. As a result, the strong reduction of $T_{\mbox{\small
C}}$ by the superexchange observed in II-VI semiconductors is
virtually absent in the case of III-V DMS.\cite{Diet00a}

Particularly interesting is the problem of interplay between
Anderson-Mott localization and carrier-mediated exchange
interaction. Our results demonstrate the presence of
magnetoresistance and temperature dependent localization,
qualitatively similar to those observed previously in n-type DMS
but somewhat enhanced due to a greater magnitude of the exchange
energy in the case of the holes. At the same time, the strength of
the carrier-mediated ferromagnetic interaction appears to be
insensitive to hole localization. This behavior, observed also in
Ga$_{1-x}$Mn$_{x}$As,\cite{Mats98} is assigned to the different
length scales involved in the two processes. However, the
low-temperature phase in nonmetallic samples show a number of
peculiarities, such as slow dynamics and partial saturation of the
magnetization. Furthermore, in contrast to, {\it e.g.}, Eu
chalcogenides, no reentrance to the metallic behavior is observed
in the low-temperature phase, which indicates that spin-disorder
scattering remains effective. Thus, the ground state on the
insulator side of the MIT may not be a simple co-linear
ferromagnet.  The presence of a phase separation into ferromagnetic
and paramagnetic spin puddles has been suggested\cite{Diet00a} to
occur near the MIT. We plan to employ local experimental probes in
order to verify this conjecture. We also hope that our results will
prompt a theoretical examination of effects of mesoscopic
fluctuations in the density-of-states and localization radius upon
the carrier-mediated ferromagnetic interactions between magnetic
impurities.

\section*{Acknowledgments}
We thank Jacek Furdyna, Hideo Ohno, and Maciej Sawicki for valuable
discussions. The Grenoble-Warsaw collaboration was supported by the
"Polonium" Program, the work in Poland by the Foundation for Polish
Science and the State Committee for Scientific Research under Grant
No. 2-PB03B-02417.

\end{document}